# Stroboscopic Tracking of a Random Walker

R. Mansilla

Centro de Investigaciones Interdisciplinarias en Ciencias y Humanidades. Universidad Nacional Autónoma de México. Torre II de Humanidades, Ciudad Universitaria, México, 04360, DF. México.
Centro de Ciencias de la Complejidad, Universidad Nacional Autónoma de México, Torre de Ingeniería, Ciudad Universitaria, México, 04360, DF. México.

## Abstract

The patterns of motion of mobile agents has received recently wide attention in the literature. There is a number of recent studies centered around the motion behavior of many agents ranging from albatrosses to human beings. Special attention has been given to the covered distances statistical distributions. In some cases, due to the lack of accurate data about the motion of the agents it has been necessary to plan very clever experiments to obtain them. These experiments try to infer the statistical properties of the agents' *real* motion from the *observed* positions in consecutive time intervals. The length of the time intervals are random variables taking their values from a previously known statistical distribution or from a distribution deduced from empirical data. The aim of this work is to demonstrate that for a Gaussian Random Walker it is, in general, impossible to recover the real motion patterns distribution from the stroboscopic observation of the agents. Moreover, it is also shown that the distances distribution strongly depends on the agents' observation time intervals. These claims are sustained by numerical experiments.

## Introduction

Recently, the motion patterns of living agents have received wide attention. The advent of new assorted devices such as global positioning systems, cell phones and other original forms of tracking using internet has opened new possibilities to study them, to set out new hypothesis or to validate their mathematical modeling. Well documented examples can be found for the case of albatrosses [1,2], pigeons [3], monkeys [4], jackals [5] and human beings [6,7,8,9]. In these studies the position of the living agents is recorded for consecutive time intervals. Normally, the intervals' length is constant but sometimes is taken from a statistical distribution calculated from the experimental data. A relevant question is how accurate is the global description of the agents' motion when inferred from these stroboscopic observations.

This paper shows that, for Gaussian Random Walker, a very restricted kind of motion, the statistical properties of the distances among observed positions depend on the statistical properties of stroboscopic time intervals. Our study is constricted to two distributions that are very frequent in the literature: The Levy Distribution and the Levy Distribution with exponential cutoff.

## Results

Two dimensional Random Walkers were analyzed. Their trajectories are a collection of independent and identical distributed pairs $\{(\theta_i, s_i), i = 1\ldots, k\}$, where $\theta_i$ is a random variable uniformly distributed in $(0, 2\pi]$ and $s_i$ is a Gaussian random variable with a probability density function:

$$p(s) = \frac{1}{\sqrt{2\pi\sigma}} e^{-\frac{1}{2}\left(\frac{s-\mu}{\sigma}\right)^2}$$

The variable $\theta_i$ represents the angle measured from the positive horizontal line of the i-th step and $s_i$ is the length of the step. Along every numerical experiment $\mu$ and $\sigma$ remain constant. Time is assumed to be discrete and during each unit time interval $\delta t$ one and only one walk step is executed.

The random walker trajectories are stroboscopically observed. The time interval among observations is $\Delta t_k \delta t$. The $\Delta t_k$ are random variables with Levy or Truncated Levy probability density functions. This choice is supported by the apparent ubiquity of these distributions [1, 2, 4, 6, 7, 8, 9] in Nature.

If $\Delta t$ denotes the time interval between two consecutive observations of the random walker, then his distribution is said to be *Levy* when $p(\Delta t) \propto \Delta t^{-\alpha}$ for some exponent $\alpha$. For example, in [6] authors comment that the distribution of rests between observed displacements of dollar bills fits this distribution with $\alpha \approx 1.6$.

On the other hand the distribution of the time intervals is said to be *Truncated Levy* when $p(\Delta t) = \Delta t^{-\alpha} e^{-\tau \Delta t}$, where $\alpha$ and $\tau$ are both constant. The cutoff is $1/\tau$. This distribution has been extensively used in the current literature. To illustrate, in [8] can be found that the distribution of the time interval $\Delta t$ between two consecutive phone calls fits this distribution with $\alpha = 0.9$ and $\tau = 1/48$ days when time intervals $\Delta t$ and the probability $p(\Delta t)$ are properly rescaled.

The core of our simulations is now described. In Fig. 1 the real trajectory of the random walker (blue line) and the stroboscopic trajectory (red line) are shown. The red dots represent the positions over the trajectory when the observations were made. In this case the time intervals between observations were taken from a Truncated Levy distribution. Fig. 2 shows the probability density function of steps in the real trajectory (blue line in Fig. 1) and Fig. 3 shows the probability density function in the observed trajectory (red line in Fig 1). They are evidently different.

Simulations were carried out for $\mu = 50, 55, 60, \ldots, 200$. For each $\mu$, values of $\sigma \in \left\{ \frac{\mu}{10}, \frac{\mu}{10} + 1, \ldots, \frac{\mu}{5} \right\}$ were used. In other words; taking the standard deviation between the 10% and 20% of $\mu$. This totals 806 pairs of parameters. For every one of these pairs $(\mu, \sigma)$ several simulations were carried out changing the parameters of the probability density functions used to generate the time intervals between consecutive observations. In every case, the number of steps in the random walkers trajectories was $k = 250000$. The probability density function was computed for all the walks both for the real and the observed trajectories. Moreover, for every case the best fitting theoretical distribution parameters were calculated.

Separate sets of experiments were carried out for Levy and Truncated Levy distributions:

**A.** Experiments with $p(\Delta t) \propto \Delta t^{-\alpha}$ distributions of time intervals $\Delta t$ are described: For each pair $(\mu, \sigma)$ simulations with $\alpha = 1, 1.1, 1.2, \ldots, 2$ were carried out. In each of those experiments a random walk was generated and the corresponding probability density function for the observed trajectory was calculated. The best fit was a Levy distribution. A random sample of those experiments is shown in Fig. 4. Each subplot is labeled **mean-X-std-Y-alpha-Z**, where X stands for the mean $\mu$ of the steps in the real trajectory, Y stands for the standard deviation $\sigma$ and Z is the $\alpha$ value in the Levy distribution from which the time intervals were taken. The red line represents the observed distances distribution while the blue line represents the Levy distribution fitted to the data. The regression was made in logarithmic scales and $R^2$ was calculated in every case. The $R^2$ distribution is shown in Fig. 5.

**B.** The numerical experiments with $p(\Delta t) \propto \Delta t^{-\alpha} e^{-\tau \Delta t}$ distributions for the time intervals $\Delta t$ were made in a similar way. For every pair $(\mu, \sigma)$ out of the 806 possibilities, simulations were made with $\tau = 1, 1.1, 1.2 \ldots, 5$ and $\alpha = 0.9$. In each one a random walk was generated and the probability density function for the observed positions was calculated. In every single case, the best fit was found when using a Truncated Levy distribution. A random sample of these experiments can be seen in Fig. 6. The same way as in the previous case, each subplot is labeled **mean-X-std-Y-tau-Z**, where X stands for the mean $\mu$ of the real trajectory steps, Y for the standard deviation $\sigma$ and Z the $\tau$ value used in the Truncated Levy distribution from which the time intervals between observations were taken. The red line represents the observed distances distribution while the blue line represents the Truncated Levy distribution fitted to the data The regression to the Truncated Levy distribution was made in logarithmic scales and the $R^2$ for every case was calculated. Its distribution is shown in Fig. 7.

An additional test showing that stroboscopic trajectories do not follow a Gaussian random walk is the quadratic mean displacement $\langle F^2(\Delta t)\rangle$ [4], [5]. It is well known that for a Gaussian random walk $\langle F^2(\Delta t)\rangle \propto \Delta t$.

The quadratic mean displacements distribution was calculated for the first analyzed case (i.e. Levy time intervals). A random sample of the experiments is shown in Fig. 8. Here the labels in the subplots are **mean-X-std-Y-alpha-Z**, where X stands for the mean $\mu$ of the real trajectory steps, Y the standard deviation $\sigma$ and Z the $\alpha$ value in the Levy distribution were the time intervals were taken. The results show that the mean square displacement scales as $\langle F^2(\Delta t)\rangle \propto \Delta t^{\alpha}$ with $\alpha = 2 \pm 0.05$. The $\alpha$ exponents distribution can be seen in Fig. 9. The regression $\langle F^2(\Delta t)\rangle \propto \Delta t^{\alpha}$ was made in logarithmic scales and $R^2$ was calculated in every case. Their distribution appears in Fig. 10. Similar results were obtained for the Truncated Levy distribution.

**Discussion**

If the agents under study behave as Gaussian random walkers, our study concludes that the distribution of the length steps between observed positions is similar to the distribution of the time interval lengths between such observations, at least for the ubiquitous Levy type or Truncated Levy distributions. However, an accurate recovery of the trajectories of the Gaussian random walker could be done with constant length interval of observation. In this case the Central Limit Theorem assures similar parameters in the distribution.

The result of this paper is a warning to the researches in this area, because in the case of a Gaussian random walker the real distribution of the length steps is not recovered from the distribution of length steps in observed positions. Therefore, how to be sure that the distribution of time intervals between observations will not affect the recovery of real distribution of steps in other scenarios?

A more general question is posed. Let $g(s)$ the probability density function of the distances and $f(\Delta t)$ the probability density function of time intervals between observations. To find sufficient conditions over $f(\Delta t)$ to recover $g(s)$ is still an open problem.

Similar result has been found with spatial restriction. The subordination of the observed distribution to the spatial properties of the environment where agents move have been already studied [10]. Results analogous to ours were found.

**Acknowledgments**

The author would like to thanks to P. Miramontes and G. Cocho for some fruitful discussions, to O. Miramontes for call our attention to references [2] and [10], to N. Del Castillo for provide very important advices in the programming tasks, and to J. L. Gordillo of DGSCA, the supercomputing facility of UNAM for all the support offered during the calculations.

**Figure Legends**

**Figure 1. Real and observed trajectories.**
The trajectory of the Gaussian random walker (blue line) and the observed trajectory (red line). The red dots represent the position where observations of real trajectory were done.

**Figure 2. Distribution of the steps in real trajectory**
The probability density function of the steps in real trajectory of the experiment described in Figure 1. See text for details

**Figure 3. Distribution of the steps in the observed trajectory**
The probability density function of the steps in observed trajectory of the experiment described in Figure 1. See text for details.

.

**Figure 4. The fitness to Levy distribution.**
A random sample of 12 numerical experiments developed taken the probability density function of the time intervals between observations as a Levy distribution. The red lines represent the distribution of length steps in the observed trajectory and the blue line the fitted Levy distribution. See text for details.

**Figure 5. The goodness of the fit.**
The distribution of the $R^2$ coefficients in the experiments shown in Fig. 4.

**Figure 6. The fitness to Levy distribution with exponential cutoff.**
A random sample of 12 numerical experiments developed taken the probability density function of the time intervals between observations as a Levy distribution with exponential cutoff. The red lines represent the distribution of length steps in the observed trajectory and the blue line the fitted Levy distribution with exponential cutoff. See text for details.

**Figure 7. The goodness of the fit.**
Distribution of the $R^2$ coefficients in the experiments shown in Fig. 6

**Figure 8. The behavior of the mean squared displacement.**

A random sample of 12 calculated $\left\langle F^2(\Delta t) \right\rangle$ in the experiments shown in Fig. 4. See text for details.

**Figure 9. The distribution of exponents.**

The distribution of the $\alpha$ exponents in the fittings $\left\langle F^2(\Delta t) \propto \Delta t^{\alpha} \right\rangle$ shown in Fig. 8.

**Figure 10. The goodness of the fit.**

The distribution of the $R^2$ in the fittings shown in Fig. 7. See the text for details.

.

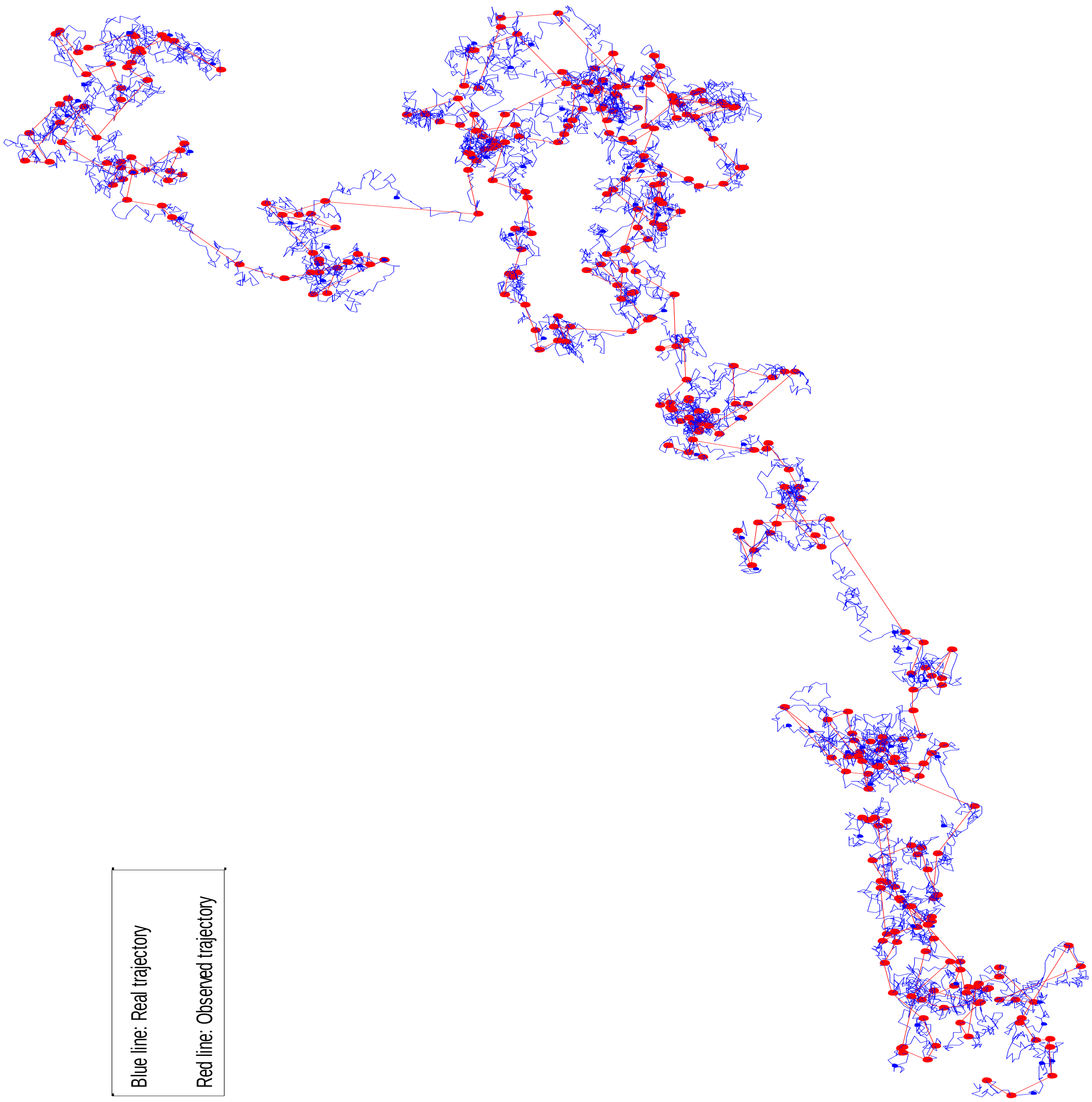

**Figure 1**

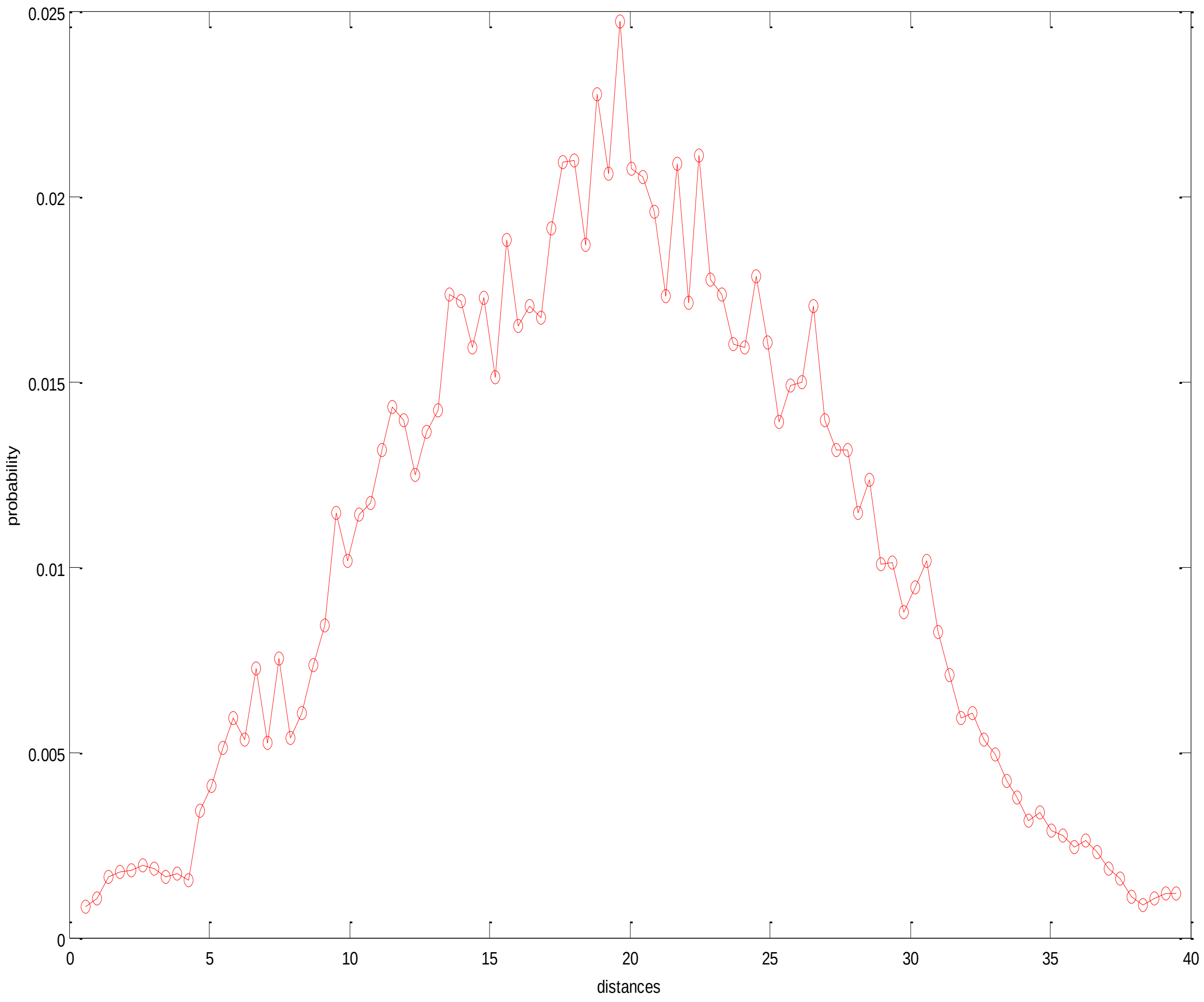

**Figure 2**

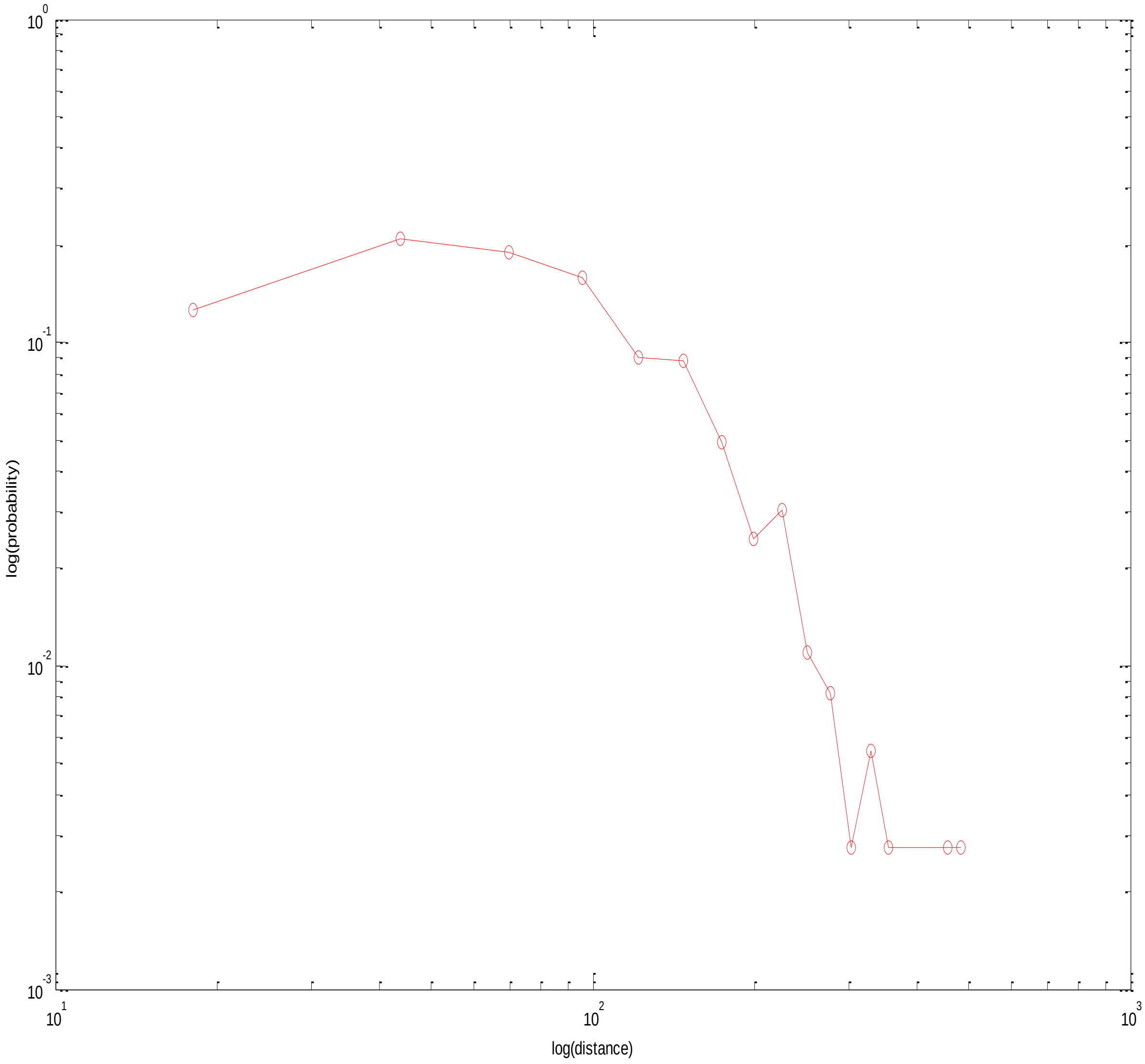

**Figure 3**

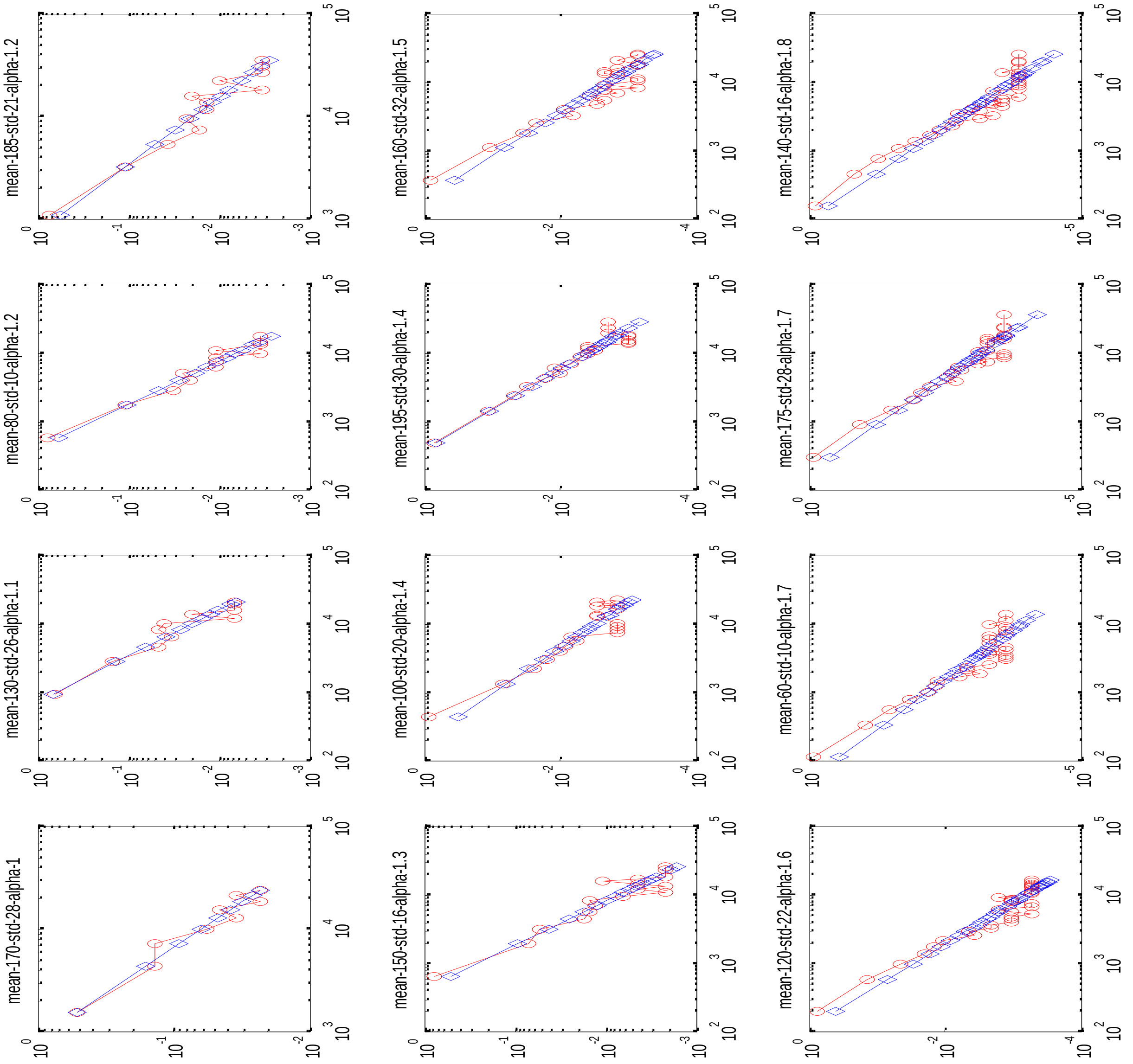

**Figure 4**

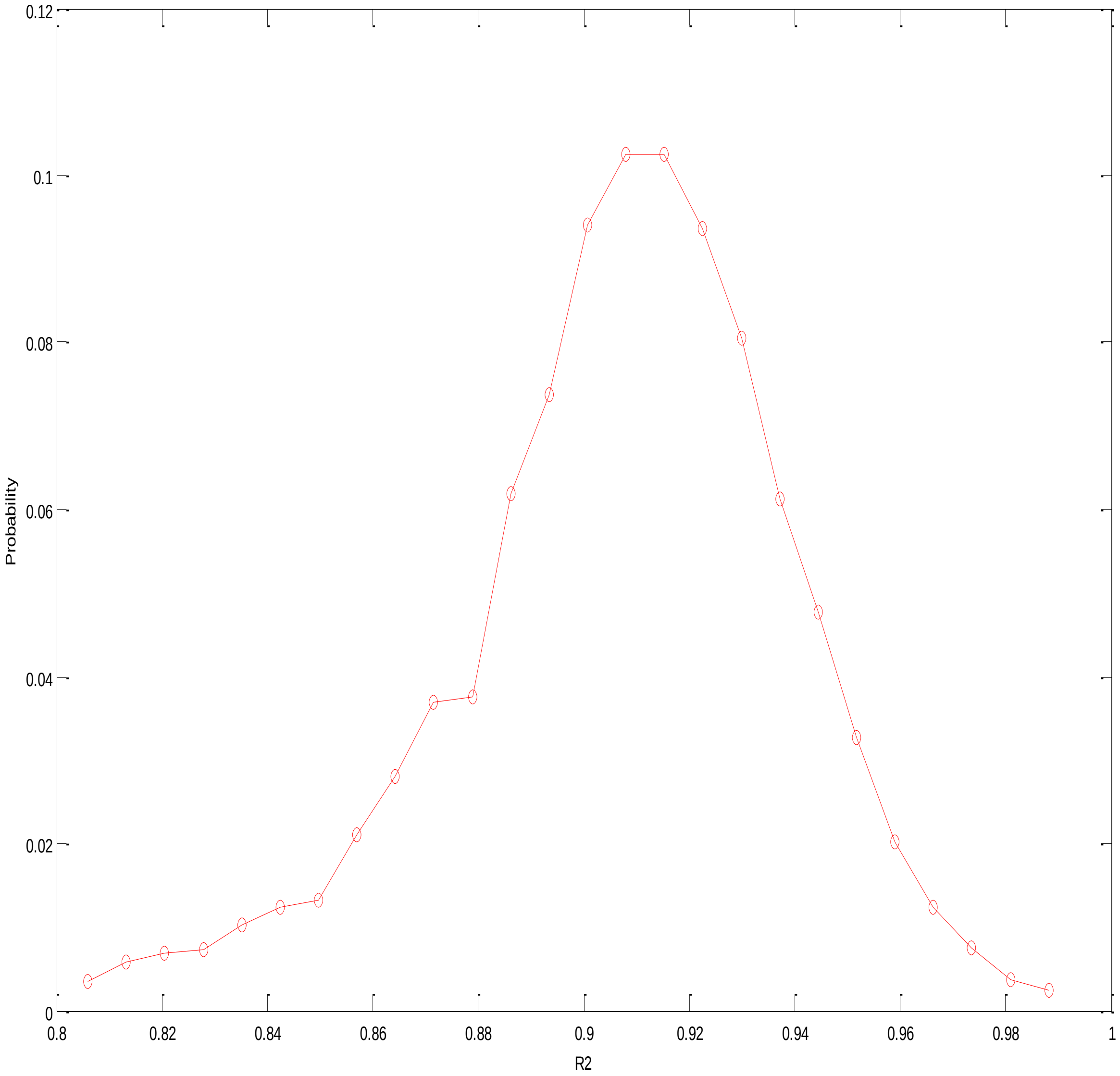

**Figure 5**

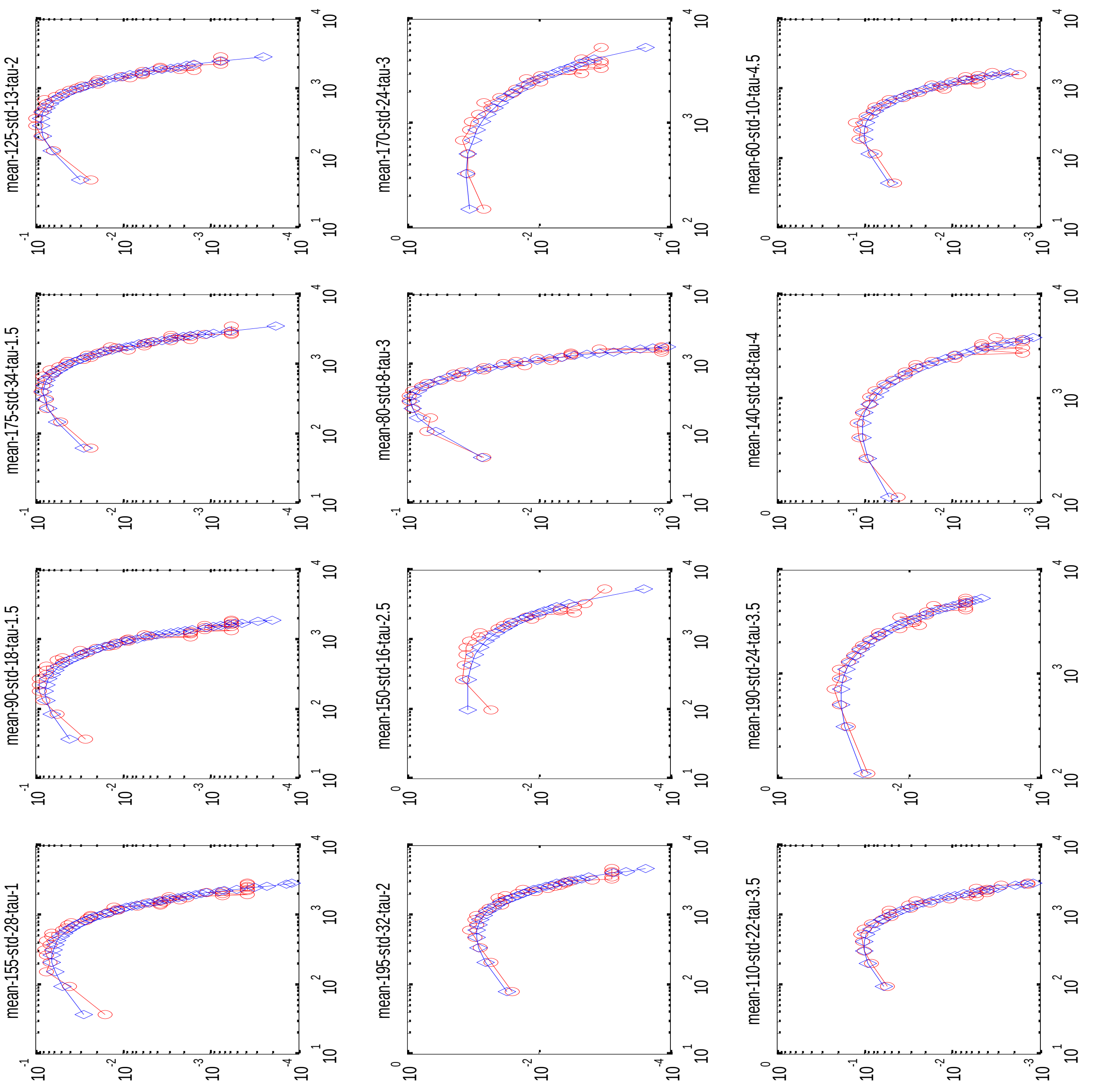

**Figure 6**

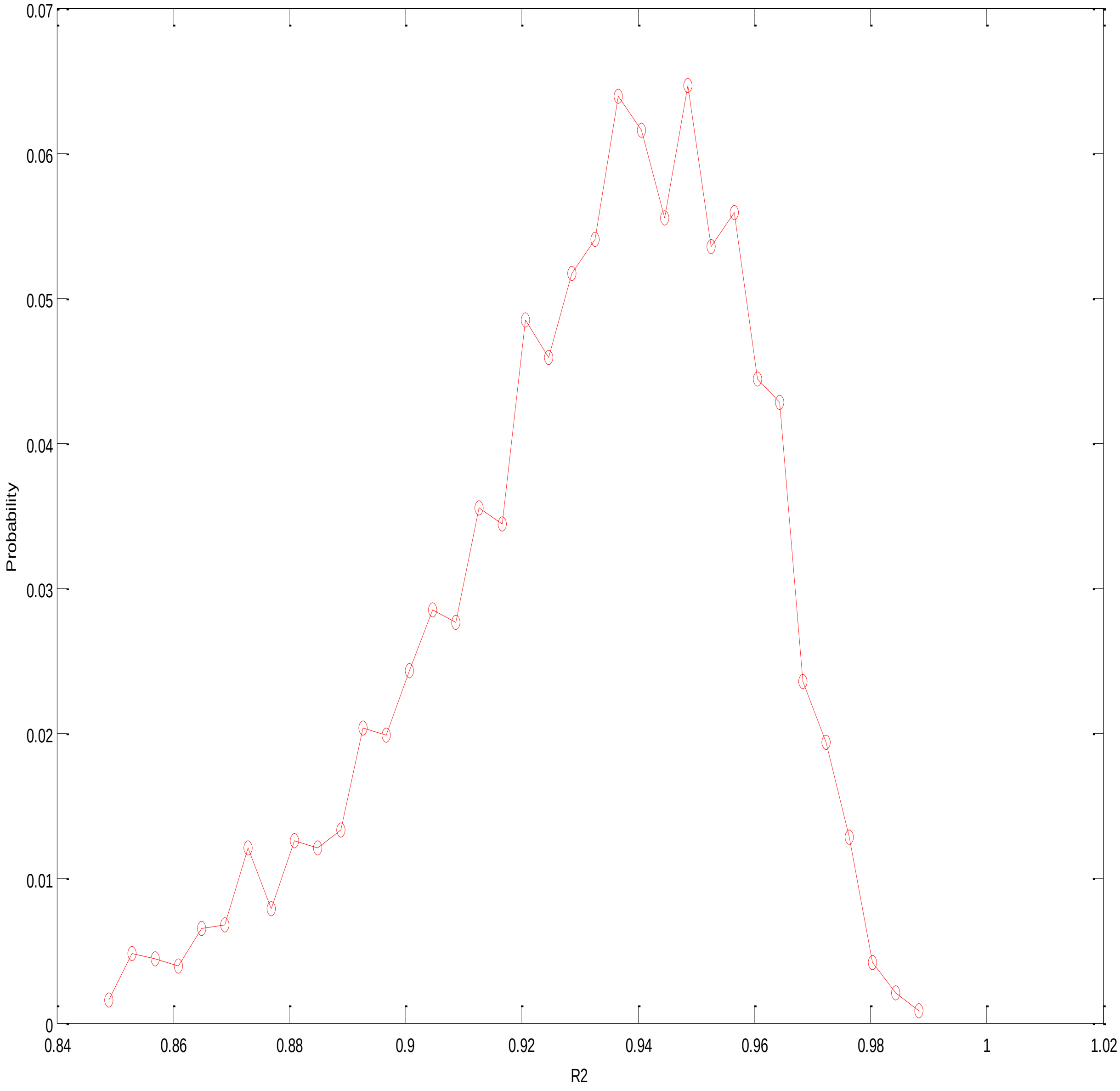

**Figure 7**

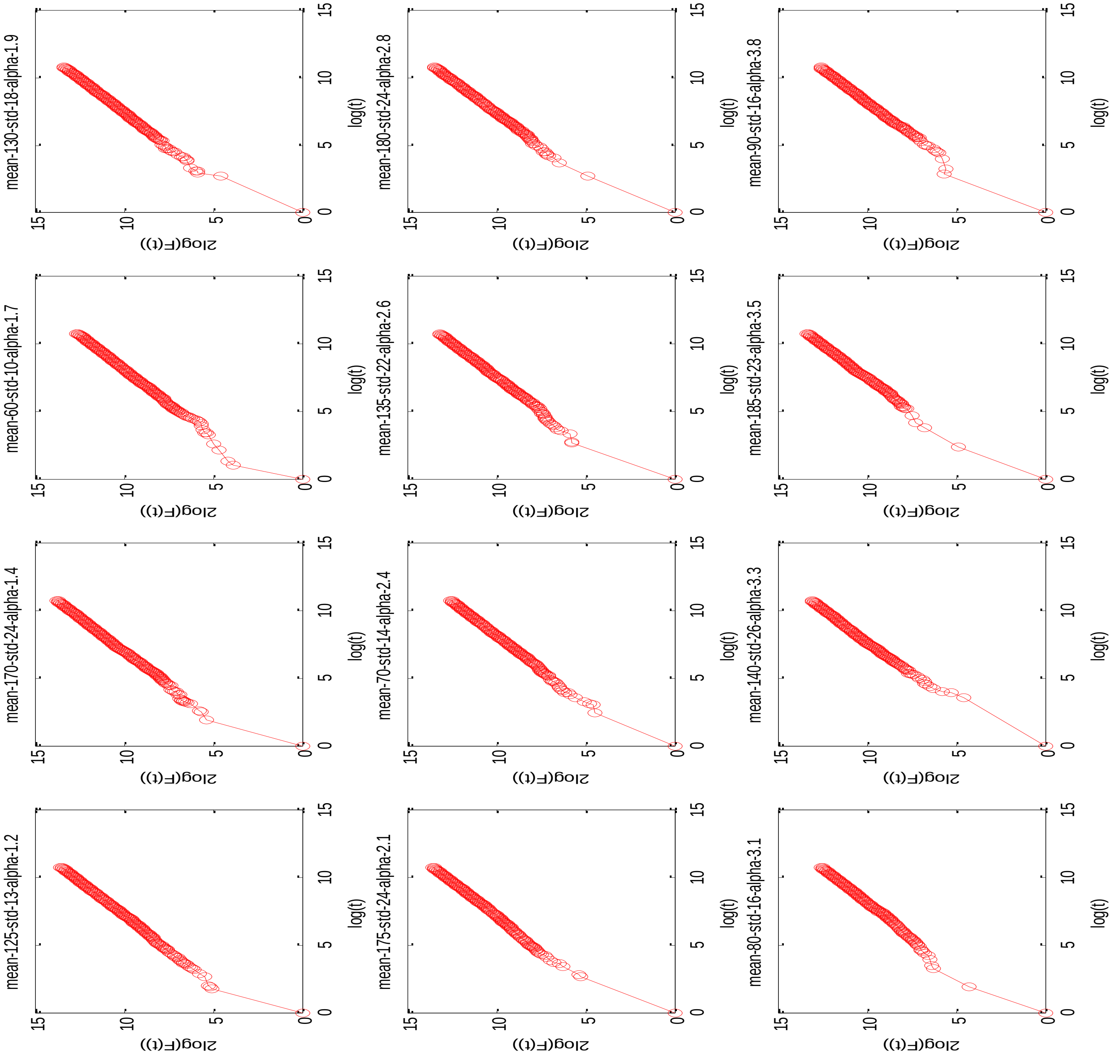

**Figure 8**

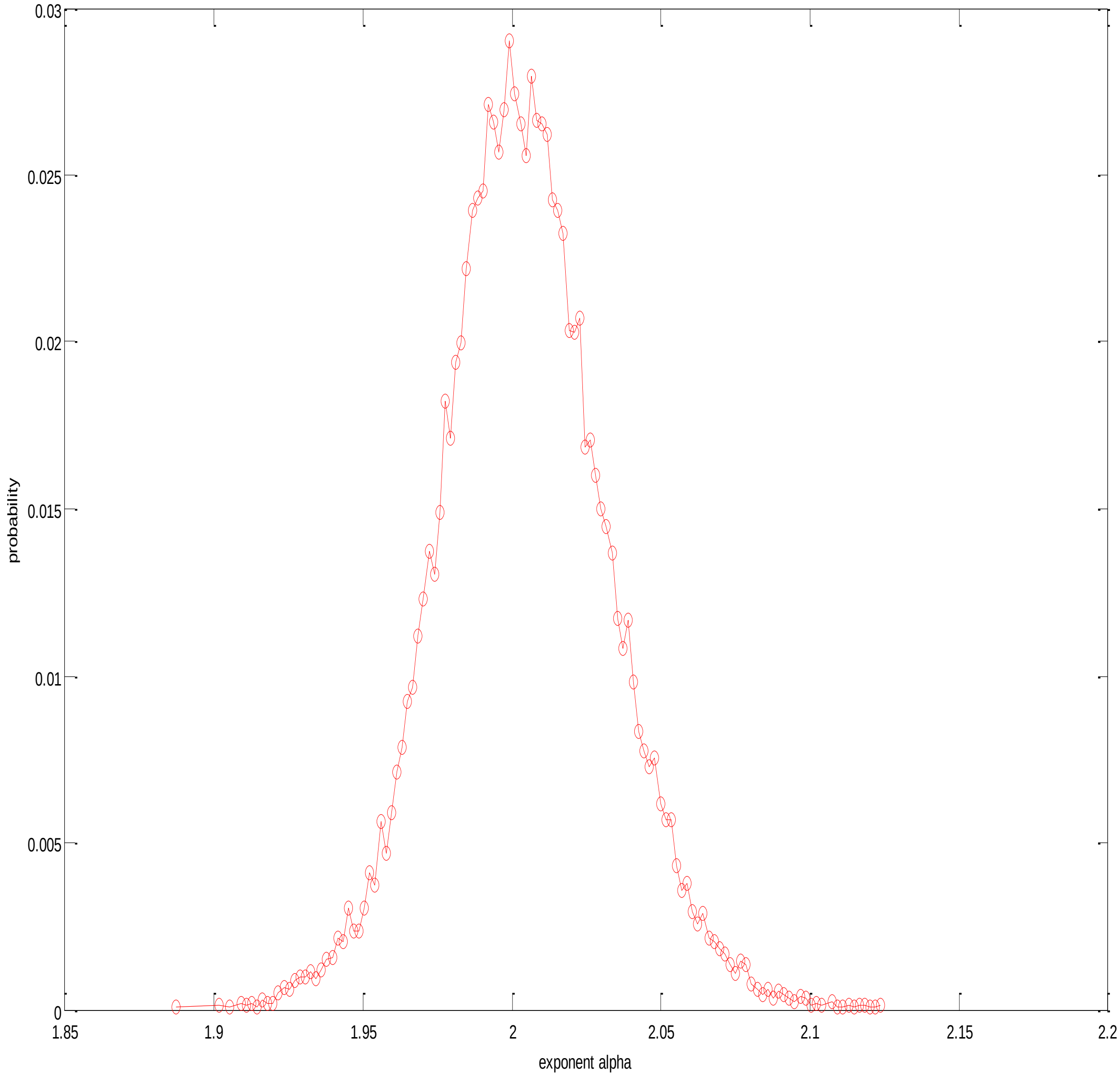

**Figure 9**

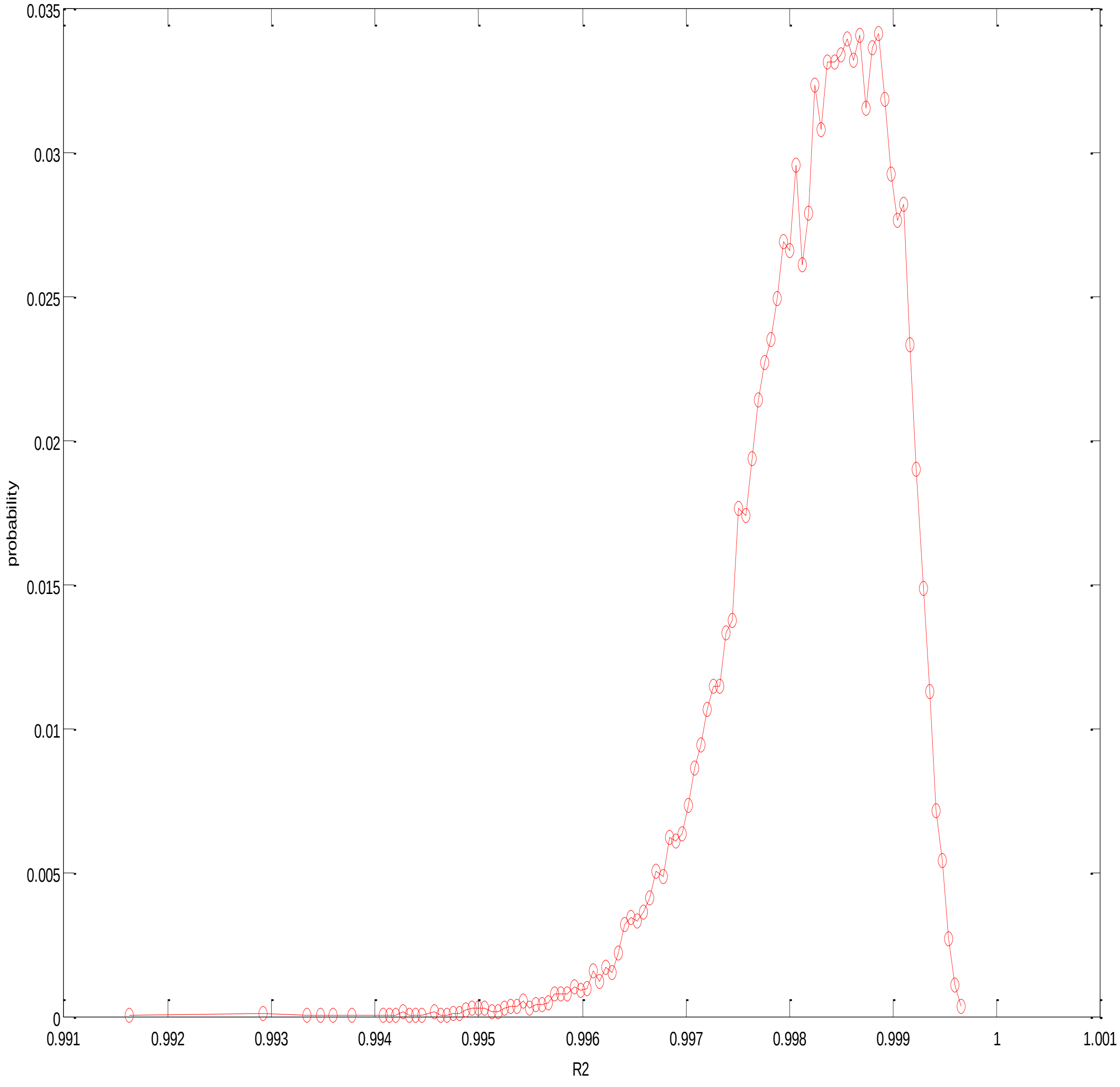

**Figure 10**